%% file: main.tex
\documentclass[10pt,twocolumn,preprintnumbers,amsmath,amssymb,nofootinbib,superscriptaddress, floatfix]{revtex4-2}

\input{packages}

\def \be {\begin{equation}}
\def \ee {\end{equation}}
\def \dd {\mathrm{d}} 
\def \t {\tilde}
\def \p {\partial}
\def \l {\left}
\def \r {\right}
 
\def \dph {\delta\ph}
\def \bph {\bar{\ph}}
\def \bs {\boldsymbol}

\def \ii {\mathrm{i}}

\def \P {\mathcal{P}}
\newcommand{\e}[1]{_{\rm #1}}
\newcommand{\FT}[1]{\tilde{#1}}
\newcommand{\ex}[1]{\mathrm{e}^{#1}}

\def \ph {\varphi}
\def \dph {\delta\ph}
\def \bph {\bar{\ph}}

\newcommand{\bG}{\bar{G}}

\newcommand{\beq}{\begin{equation}}
\newcommand{\eeq}{\end{equation}}
\newcommand{\bea}{\begin{eqnarray}}
\newcommand{\eea}{\end{eqnarray}}


\newcommand\ees{\end{eqnarray}}
\newcommand\bees{\begin{eqnarray}}

\definecolor{magenta}{rgb}{0.1,0.98,0.6}
\definecolor{dgreen}{rgb}{0,0.7,0.0}

\begin{document}
\title{Scalar \texorpdfstring{$\check{\hbox{C}}$}{Ch}erenkov radiation from high-energy cosmic rays}

\author{Charles Dalang}
\email{charles.dalang@unige.ch}
\affiliation{Universit\'e de Gen\`eve, D\'epartement de Physique Th\'eorique and Center for Astroparticle Physics, 24 quai Ernest-Ansermet, CH-1211 Gen\`eve 4, Switzerland}

\author{Pierre Fleury}
\email{pierre.fleury@uam.es}
\affiliation{Instituto de F\'isica Te\'orica UAM-CSIC,
Universidad Aut\'onoma de Madrid,\\
Cantoblanco, 28049 Madrid, Spain}

\author{Lucas Lombriser}
\email{lucas.lombriser@unige.ch}
\affiliation{Universit\'e de Gen\`eve, D\'epartement de Physique Th\'eorique and Center for Astroparticle Physics, 24 quai Ernest-Ansermet, CH-1211 Gen\`eve 4, Switzerland}

\begin{abstract}
As first noted by Robert Wagoner in the nineteen-seventies, if a scalar field is nonminimally coupled to the Ricci scalar and propagates at subluminal speeds, then there exists the possibility of scalar $\check{\hbox{C}}$erenkov radiation from a moving particle. The mere observation of high-energy cosmic rays could in principle rule out the existence of such scalar fields since any particle moving faster than scalar perturbations would lose energy in the form of scalar waves until it moves slower than those. We compute in detail the energy loss to scalar waves and find that it scales with the square of the ultra-violet (UV) cutoff frequency of the effective field theory (EFT) of gravity. For dark-energy-motivated EFTs, the UV cutoff can be low, in which case that energy loss could always be negligible. In contrast, if viewed as a covariant theory valid at all scales or as an EFT valid at higher energies, perhaps even all the way up to the Planck scale, as may be the case if motivated by quantum-gravity perspectives, then the energy loss to scalar waves may diverge or become dramatically large. In this case, high-energy cosmic rays of extragalactic origin stringently constrain any conformally coupled scalar fields with non-canonical kinetic terms, although a minimum scalar phase velocity is required to trust the EFT. 
\end{abstract}

\preprint{IFT-UAM/CSIC-21-100}
\maketitle

\section{Introduction}\label{sec:intro}

\lettrine{T}{here} exists, in the context of gravitational physics, an ambiguity between missing matter and describing the Universe with an incomplete theory of gravity. As an example, unexplained perturbations of the orbit of Uranus led the French astronomer A.~Bouvard to predict, in 1821, the existence of a missing planet, known as Neptune, which was observed and identified as such for the first time more than two decades later by J.~C.~Adams, in 1843 \cite{JCAdams:1946}. Conversely, it was working with the inadequate theory of Newtonian gravity and the observation of the perihelion advance of Mercury that led U.~Le Verrier to think of a missing unobserved planet, named Vulcan, which this time, was never discovered~\cite{Vulcan}. The theory of gravity was incomplete and a more fundamental theory of gravity, general relativity, was shown to solve this missing-matter issue. With this historical perspective in mind, and the contemporary problems of cosmology, such as the inferred missing dark matter and mysterious dark energy when adopting general relativity, it seems reasonable to question the currently accepted theory of gravitation.

Alternative theories of gravity of present interest include scalar-tensor theories and their generalizations, including Horndeski theories~\cite{Horndeski:1974wa} and beyond~\cite{Gleyzes:2014dya,BenAchour:2016fzp,Kobayashi:2011nu,Jana:2020vov}. A subset of those are able to affect cosmological dynamics and evade, via so-called screening mechanisms \cite{Vainshtein:1972sx,Khoury:2003aq,Babichev:2009ee}, tight precision tests of gravity which have been conducted within the Solar System \cite{Hofmann:2018myc, Bertotti:2003rm} and within galactic environments with binary pulsars \cite{PhysRevD.45.1840,Kramer:2006nb}, for example. A powerful test of those theories, which circumvents the problem of screening, comes from the measurement of the speed of gravitational waves (GWs) through cosmological distances, which practically eliminates a large parameter space of gravitational modifications for effective field theories of modified gravity and dark energy~\cite{McManus:2016kxu, Lombriser:2015sxa, Lombriser:2016yzn, Ezquiaga:2017ekz, Creminelli:2017sry}. This constraint arises because many of these scalar-tensor interactions imply non-luminal propagation of GWs due to the nontrivial time dependence of the background scalar field configuration. Another, perhaps even more exciting consequence of the observation of gravitational waves, is that those do not seem to decay into scalar waves, as predicted by degenerate higher-order scalar-tensor (DHOST) theories, thereby eliminating their cosmological viability~\cite{Creminelli_2018}.

As a follow-up question and since those constraints have been so powerful, can we learn anything about gravitational theories from the propagation of other fields such as high-energy cosmic particles? It turns out that for metric theories which possess a direct coupling of the scalar field to the Ricci scalar, scalar perturbations are sourced by the trace of the energy-momentum tensor of matter fields. Their equation of motion is reminiscent of the electromagnetic potential which is sourced by electrically charged particles that lose energy to electromagnetic waves if they travel faster than the speed of light in a medium. This emitted electromagnetic signal, known as $\check{\hbox{C}}$erenkov radiation~\cite{PhysRev.52.378} compensates the energy loss per unit distance of those particles, as predicted by I.~Franck and I.~Tamm \cite{Frank:1937fk}, leading to their Nobel Prize shared with P.~$\check{\hbox{C}}$erenkov in 1958. Tight constraints on the speed of spin-2 GWs via GW $\check{\hbox{C}}$erenkov radiation have already been inferred from the observation of high-energy particles~\cite{Kimura:2011qn}. An obvious follow-up question is then whether high-energy particles lose their energy by emitting scalar waves if they travel faster than those~\cite{PhysRevD.1.3209,Elliott:2005va}. After all, high-energy particles are observed all the time. Could it be that their mere observation rules out a significant class of scalar-tensor theories? 

In Sec.\,\ref{sec:Scalar_Cherenkov_Radiation}, we explore for the first time the energy loss per unit distance of a massive particle to scalar waves due to it traveling faster than the speed of scalar waves in a conformally coupled k-essence theory \cite{Armendariz-Picon:1999hyi,Armendariz-Picon:2000ulo}. In Sec.\,\ref{sec:discussion}, we discuss constraints on the ultra-violet (UV) cutoff of the effective field theory (EFT) of gravity for conformally coupled k-essence models and also discuss more generic phenomenological models. Finally, we present conclusions in Sec.\,\ref{sec:conclusions}. 

We adopt the $(-,+,+,+)$ signature for the metric tensor, Greek indices run from $0$ to $3$ and Latin indices from $1$ to $3$. A comma indicates a partial derivative, $Z_{,\mu}\equiv \partial_\mu Z$, while a semicolon denotes a covariant derivative associated with the Levi-Civita connection, $Z_{\mu;\nu}\equiv \nabla_\nu Z_\mu$. Bold symbols represent Euclidean three-vectors. A bar indicates a background quantity. Units are such that $c=\hbar=1$.

\section{Scalar \texorpdfstring{$\check{\hbox{C}}$}{Ch}erenkov Radiation}\label{sec:Scalar_Cherenkov_Radiation}
In this section, we derive an equivalent of the Franck-Tamm formula for scalar $\check{\hbox{C}}$erenkov radiation.  We start by considering a generic conformally coupled k-essence theory, which is sufficient for massive particles to source scalar perturbations. The k-essence contribution is needed to modify the phase velocity of the scalar field propagating in a homogeneous and isotropic background cosmology.

\subsection{Action and equation of motions}
We consider conformally coupled k-essence theories described by the following Jordan-frame action
\begin{align}\label{eq:Action}
 S=\frac{M\e{P}^2}{2} \int_\mathcal{M}\dd^4 x\sqrt{-g} \l[G_4(\varphi) R + G_{2}(\varphi, X) \r] + S_{\rm m}[\Psi_{\rm m},g_{\mu\nu}]\,,
\end{align}
where $M\e{P}\equiv 1/\sqrt{8\pi G}$ denotes the reduced Planck mass, $g=\det(g_{\mu\nu})$ is the determinant of the spacetime metric $g_{\mu\nu}$, $R$ denotes the Ricci scalar and $G_2(\varphi,X)$ and $G_4(\varphi)$ are free functions of the scalar field $\ph$ and its canonical kinetic term $X\equiv -\p_\mu \ph \p^\mu \ph/2$. We chose to work with a dimensionless scalar field $\varphi$. The matter action $S_{\rm m}$ depends on the matter fields $\Psi_{\rm m}$ and on the spacetime metric $g_{\mu\nu}$ but not on $\ph$. We study linear perturbations of the metric and of the scalar field
\begin{align}
g_{\mu\nu}& = \bar{g}_{\mu\nu}+h_{\mu\nu}\,, \qquad ||h_{\mu\nu}||\ll 1\,,\\
\varphi & = \bph + \dph\,,\qquad \qquad ~ |\dph|\ll 1\,,
\end{align}
propagating through arbitrary backgrounds for the metric $\bar{g}_{\mu\nu}$ and for the scalar field $\bph$, which may be spacetime dependent. For sufficiently high-frequency waves,\footnote{By high-frequency waves, we mean waves whose wavelength is much smaller than the scale on which the background spacetime metric varies appreciably. This condition will translate into an infrared cutoff $\Omega\e{IR}$.} the equations of motions may be written as~\cite{Dalang:2020eaj}
\begin{align}
\slashed{K}\indices{_\varphi ^\varphi ^\alpha ^\beta} \dph_{;\alpha\beta} + \mathcal{O}(\dph_{;\alpha})= \frac{\bG_{4,\ph}}{\bG_4}M\e{P}^{-2} \delta T\,,
\end{align}
where $\slashed{K}\indices{_\varphi ^\varphi ^\alpha ^\beta}$ is a kinetic rank-2 tensor introduced in \cite{Dalang:2020eaj}, which depends on the background functions, including $\bG_2$ and $\bG_4$ and their derivatives. We neglect first-order derivatives acting on $\dph$, which only affect the amplitude of scalar waves across cosmological distances, and masslike terms, which are negligible for sufficiently high-frequency waves~\cite{Dalang:2020eaj}. Specializing to a flat homogeneous and isotropic Universe, there exists a (comoving) coordinate system $(\tau, X, Y, Z)$ in which the background metric takes on the form of a Friedmann-Lemaître metric described by the following line element
\begin{align}
\dd s^2 =  -\dd \tau^2 + a^2(\tau)\dd\bs{X}^2 \, ,
\end{align}
where $a(\tau)$ indicates the scale factor parameterized by cosmic time~$\tau$. The scalar equation of motion can then be written as a wave equation, sourced by perturbations of the trace of the energy-momentum tensor
\begin{empheq}[box=\fbox]{align}\label{eq:EoM}
\p_\tau^2\delta \ph - \l(\frac{c\e{s}}{a}\r)^2 \, \nabla_X^2 \dph = -q\e{s}\delta T 
\end{empheq}
where the squared scalar phase velocity and the effective scalar charge read
\begin{align}
c\e{s}^2(\tau) & =\frac{\bG_{2,X} +3 \bG_{4,\ph}^2 \bG_4^{-1} }{\bG_{2,X} +3 \bG_{4,\ph}^2 \bG_4^{-1}  + \bG_{2,XX} \dot{\bph}^2}\,, \label{eq:cs2} \\
q\e{s}(\tau) & =\frac{\bG_{4,\ph} \bG_4^{-1} M\e{P}^{-2}}{\bG_{2,X} +3 \bG_{4,\ph}^2 \bG_4^{-1}  +  \bG_{2,XX} \dot{\bph}^2} \,,\label{eq:qs}
\end{align}
where a dot indicates a derivative with respect to $\tau$. If the equation of motion~\eqref{eq:EoM} seems to propagate scalar waves of all frequencies at the same phase velocity, it is in fact not the case. For waves with a frequency approaching the UV cutoff~$\Omega\e{UV}$ of the EFT, the irrelevant operators neglected in Eq.~\eqref{eq:Action} kick in, bringing back the phase velocity of scalar waves to the speed of light if the theory admits a Lorentz-invariant high-energy UV completion~\cite{deRham:2018red}. 

\subsection{Matter source}

\begin{figure}
    \centering
    \includegraphics[scale=0.8]{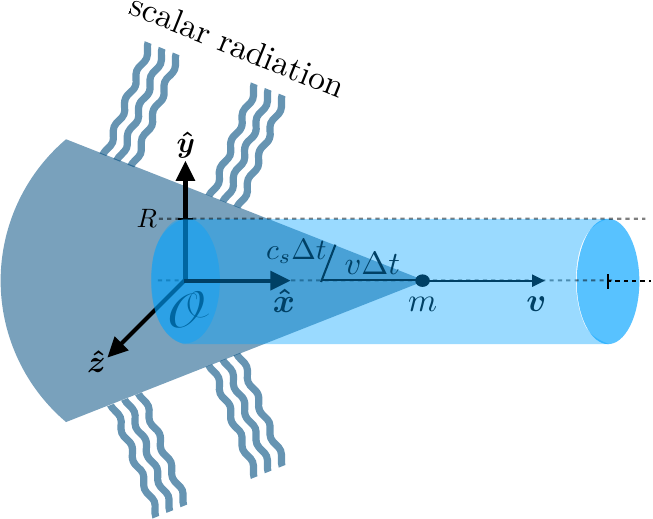}
    \caption{The system configuration for which we compute the emission of scalar $\check{\hbox{C}}$erenkov radiation. A point-like particle moves along the $x$-axis with velocity $\bs{v}$. In analogy to electromagnetic $\check{\hbox{C}}$erenkov radiation, the cone of scalar $\check{\hbox{C}}$erenkov radiation is plotted and we integrate the energy flux through the closed surface of an infinite cylinder of radius $R$, centered on the particle's trajectory.}
    \label{fig:systemconfiguration}
\end{figure}

In the previous section, we have shown how a conformal coupling leads matter fields to source scalar waves through the trace of their energy-momentum tensor. From now on, we work on scales much smaller than the Hubble radius, and use Riemann normal coordinates~$(t,x,y,z)$ that span the tangent space of the particle at some initial time, and such that $t$ locally coincides with cosmic time. We consider scalar waves sourced by a relativistic particle of mass $m$ moving with constant velocity $v$ along the $x$ axis, as depicted in Fig.~\ref{fig:systemconfiguration}. Its four-momentum reads $p^\mu =m u^\mu = (\gamma m, \gamma m \bs{v})$, where the relativistic Lorentz factor is $\gamma=1/\sqrt{1-v^2}$. Its relativistic energy-momentum tensor takes on the following form \cite{Fleury:2019ter}
\begin{align}
\delta T^{\alpha\beta}  = \frac{p^\alpha p^\beta}{p^0} \, \delta^{(3)}[\bs{x}-\bs{x}\e{p}(t)]\,,
\end{align}
where $\bs{x}\e{p}(t)$ indicates the time-dependent spatial position of the particle. In the Riemann coordinate system that we adopted,
\begin{align}
\eta_{\alpha\beta}p^\alpha p^\beta = - m^2 \,.
\end{align}
Hence,
\begin{align}\label{eq:EMTparticle}
\delta T = \eta_{\alpha \beta} \, \delta T^{\alpha \beta}
= -\frac{m}{\gamma} \, \delta(x- v t)\delta(y)\delta(z)\,,
\end{align}
where we have centered the coordinate system on the particle's trajectory.

\subsection{Radiated energy}

To compute the energy that is radiated away in the form of scalar waves by a moving particle, we need the energy-momentum tensor\footnote{This energy-momentum tensor is different from the source of the scalar waves in Eq.\,\eqref{eq:EoM}. It serves to compute the energy that is radiated away in the form of $\dph$-waves sourced by a massive particle with the trace of its EMT given by $\delta T$.} of gravitational waves in this conformally coupled k-essence theory. Expanding the tensor equation of motion to second order in $\dph$ and $h_{\mu\nu}$, expressing everything in terms of the pure spin-2 perturbation~$\gamma_{\mu\nu}\equiv h_{\mu\nu}-(h/2)\bar{g}_{\mu\nu}-({\bG_{4,\ph}}/{\bG_4}) \dph$ and the spin-0 perturbation $\dph$, time averaging\footnote{This time average should be performed on a timescale that is much larger than the period of the gravitational wave, but smaller than timescales over which the background curvature evolves appreciably.} in coordinates such that $h= 0 = \gamma\indices{^\mu^\nu _; _\nu}$, after several integration by parts\footnote{The fact that away from the source, $\gamma_{\mu\nu}$ and $\dph$ are solutions of vacuum wave equations implies that the solutions are of the form $\dph=\dph(\bs{k}\cdot\bs{x}-\omega t)$, which allows to integrate by part the spatial derivatives, in addition to the temporal ones \cite{Maggiore:2007ulw}.} and using the equations of motion, we get
\begin{align}\label{eq:GW_EMT}
T_{\mu\nu}^{\e{GW}} = \frac{M\e{P}^2}{2} \left \langle \frac{1}{2} \bG_4 \gamma_{\rho \sigma,\mu} \gamma\indices{^\rho ^\sigma _, _\nu} + \l(\bG_{2,X} + \frac{3 \bG_{4,\ph}^2}{\bG_4} \r) \dph_{,\mu}\dph_{,\nu} \right \rangle .
\end{align}
It can be easily shown from Eq.~\eqref{eq:GW_EMT}, that wavelike spin-2 solutions\footnote{Spin-0 wavelike solutions $\dph=\dph(x-c\e{s}t)$ may have a non-zero EMT if $c\e{s}\neq c$, but fail to travel faster than themselves, which turns out to be a necessary condition to emit scalar $\check{\hbox{C}}$erenkov radiation.} with $\gamma_{\mu\nu} =\gamma_{\mu\nu}(x- t)$ have a traceless EMT in flat spacetime, preventing them from sourcing scalar $\check{\hbox{C}}$erenkov radiation through Eq.~\eqref{eq:EoM}. 

We define the radiated four-momentum density with respect to an observer with four-velocity $u^\mu$ as
\be
P\e{GW}^\nu \equiv - u_\mu T^{\mu\nu}\e{GW}\,.
\ee
Projecting onto the local space of the observer gives the energy flux density in the observer's rest frame
\be
\Pi\e{GW}^\mu = (u^\mu u_\nu + \delta^\mu_\nu)P\e{GW}^\nu\,.
\ee
For an observer at rest $u^\mu=(1,\bs{0})$, the only nonzero component comes from $T^{0i}\e{GW}$ which contains the energy flux density in the direction $i$
\begin{align}\label{eq:fluxdensity}
\Pi^i\e{GW} & =  P^{i}\e{GW} = -u_0 T^{0i}\e{GW} \\
& =\P \left \langle \dph_{,0}\dph_{,i} \right \rangle\,,
\end{align}
where
\begin{align}\label{eq:P}
\P= \frac{M\e{P}^2}{2} \l( \bG_{2,X} + \frac{3\bG_{4,\ph}^2}{\bG_4}\r)
\end{align}
has the dimension of power. The energy radiated per unit time by the particle is equal to the flux of the gravitational Poynting vector~$\bs{\Pi}\e{GW}$ through any closed surface~$\Sigma$ surrounding the particle,
\begin{equation}
\frac{\dd E}{\dd t} = \int_\Sigma \dd^2\bs{x} \; \hat{\bs{n}}\cdot\bs{\Pi}\e{GW}(t, \bs{x}) \,,
\end{equation}
where $\hat{\bs{n}}$ indicates an outgoing unit normal vector to the surface $\Sigma$. The geometry of the problem suggests to choose $\Sigma$ as an infinite cylinder about the $x$-axis and with arbitrary radius~$R$. Using cylindrical coordinates $(x, r, \phi)$ about the $x$-axis, we thus have
\begin{equation}
\frac{\dd E}{\dd t}
= 2\pi R \int_{\mathbb{R}} \dd x \; \Pi^r\e{GW}(t, x, r=R) \, .
\end{equation}

In the remainder of this calculation, we will neglect the backreaction of radiation on the particle's velocity. This assumption may seem at odds with the very purpose of the calculation, which is precisely to estimate how much a cosmic ray is slown down by scalar $\check{\hbox{C}}$erenkov radiation. Such an apparent contradiction may be viewed as an adiabatic approximation. Namely, we will perform the computation of the radiated power in the stationary regime, i.e., assuming that the particle has a constant velocity, and then use the result to compute the backreaction on the particle's velocity. In that context, the problem is stationary in the particle's rest frame, and hence
\begin{equation}
\label{eq:symmetry_WKB}
\dph(t, x, r, \phi) = \dph\left(t-\frac{x}{v}, 0, r, \phi\right) .
\end{equation}

The invariance property~\eqref{eq:symmetry_WKB}, which equally applies to $\bs{\Pi}$, allows us to proceed with the calculation of the radiated power,
\begin{align}
\frac{\dd E}{\dd t}
&= 2\pi R v \int_\mathbb{R} \dd t \; \Pi^r\e{GW}(t, x=0, r=R)
\\
\label{eq:calculation_dEdt_with_average}
&= 2\pi R v \P \int_\mathbb{R} \dd t
    \left\langle \dph_{,0}(t, R) \dph_{,r}(t, R) \right\rangle
\\
\label{eq:calculation_dEdt_without_average}
&= 2\pi R v \P \int_\mathbb{R} \dd t \; \dph_{,0}(t, R) \dph_{,r}(t, R) \, ,
\end{align}

where from Eq.~\eqref{eq:calculation_dEdt_with_average} on, we use ``$R$'' as a short-hand notation for $x=0, r=R$. In order to go from Eq.~\eqref{eq:calculation_dEdt_with_average} to Eq.~\eqref{eq:calculation_dEdt_without_average}, we used that an integral over time and a time-averaging are redundant. The remainder of the calculation is quite similar to the standard electromagnetic case~\cite{Jackson:1998nia}; it consists in rewriting Eq.~\eqref{eq:calculation_dEdt_without_average} in Fourier space as
\begin{equation}c
\label{eq:dEdt_Fourier}
\frac{\dd E}{\dd t}
=
R v \P \int_\mathbb{R} \dd \omega \;
\widehat{\dph_{,0}}^*(\omega, R) \,
\widehat{\dph_{,r}}(\omega, R) \ ,
\end{equation}
using the Plancherel-Parseval theorem, and where a hat denotes a temporal Fourier transform with the convention
\begin{align}
\hat{f}(\omega, \bs{x})
&= \int_{\mathbb{R}} \dd t \; \ex{\ii\omega t} \, f(t, \bs{x}) \, ,
\\
f(t, \bs{x})
&= \int_{\mathbb{R}} \frac{\dd\omega}{2\pi} \; \ex{-\ii\omega t} \, \hat{f}(\omega, \bs{x}) \, .
\end{align}

\subsection{Fourier-space solutions}

We now compute the Fourier transforms of $\dph_{,0}$ and $\dph_{,r}$ that are involved in Eq.~\eqref{eq:dEdt_Fourier}. For that purpose we must solve the wave equation for $\dph$
\begin{align}
\label{eq:inhomogeneous_wave_equation}
\p_t^2 \dph - c\e{s}^2 \nabla_x^2 \dph
= -q\e{s}\delta T \equiv S(t,\bs{x})\,,
\end{align}
which is now written in Riemann coordinates (whence the absence of the scale factor). The source term~$S(t,\bs{x})$ depends on $\delta T$, which is given by Eq.~\eqref{eq:EMTparticle}, so that
\begin{equation}
S(t,\bs{x})
= \frac{m q\e{s}}{\gamma} \frac{\delta(r)}{2\pi r} \; \delta(x-vt)
\end{equation}
in cylindrical coordinates. We then perform a \emph{four-dimensional} Fourier transform of Eq.~\eqref{eq:inhomogeneous_wave_equation} with the following convention
\begin{align}
\FT{f}(\omega, \bs{k})
&= \int_{\mathbb{R}} \dd t \int_{\mathbb{R}^3} \dd^3\bs{x} \;
    \ex{\ii(\omega t - \bs{k}\cdot\bs{x})} \, f(t, \bs{x}) \ ,
\\
f(t, \bs{x})
&= \int_{\mathbb{R}} \frac{\dd\omega}{2\pi}
    \int_{\mathbb{R}^3} \frac{\dd^3\bs{k}}{(2\pi)^3} \;
    \ex{\ii(\bs{k}\cdot\bs{x} - \omega t)} \, \FT{f}(\omega, \bs{k}) \ ,
\end{align}
which yields, for the source term,
\begin{align}
\t{S}(\omega,\bs{k})
&= \int_{\mathbb{R}} \dd t \int_{\mathbb{R}^3} \dd^3\bs{x} \;
    S(t,\bs{x}) \, \ex{\ii(\omega t - \bs{k}\cdot \bs{x})}\\
&= \frac{m q\e{s} }{\gamma}\, 2 \pi \delta(\omega - k_x v)
\end{align}
and eventually implies
\begin{equation}
\delta\FT{\ph}(\omega, k_x, k_r)
= \frac{m q\e{s}}{\gamma}
    \frac{2\pi\,\delta(\omega - k_x v)}{c\e{s}^2(k_x^2+k_r^2) - \omega^2} \ .
\end{equation}

In order to compute $\widehat{\dph_{,0}}$, we must take the spatial inverse Fourier transform of $\delta\FT{\ph}$, namely
\begin{align}
\widehat{\dph_{,0}}(\omega, \bs{x})
&= -\ii\omega \, \widehat{\dph}(\omega, \bs{x})
\\
&= -\ii\omega \int_{\mathbb{R}^3} \frac{\dd^3\bs{k}}{(2\pi)^3} \;
    \ex{\ii\bs{k}\cdot\bs{x}}
    \delta\tilde{\ph}(\omega, \bs{k}) \ .
\end{align}
We then use cylindrical components $(k_x, k_r, \psi)$ for $\bs{k}$, where $\psi$ denotes the polar angle that the projection of $\bs{k}$ in the $Oyz$ plan makes with $\bs{x}=(0,r,\phi)$. The differential element reads $\dd^3\bs{k}=\dd k_x (k_r \dd k_r) \dd\psi$ and the calculation goes as follows,
\begin{align}
\widehat{\dph_{,0}}(\omega, R)
&= -\frac{\ii\omega m q\e{s} }{\gamma}
    \int_{\mathbb{R}} \dd k_x \; \delta(\omega - k_x v)
    \nonumber\\ &\quad\times
    \int_0^{2\pi} \frac{\dd\psi}{2\pi} \; \ex{\ii k_r R \cos\psi}
    \nonumber\\ &\quad\times
    \int_0^\infty \frac{k_r \dd k_r}{2\pi} \;
    \frac{1}{c\e{s}^2(k_x^2+k_r^2) - \omega^2} 
\\
&= -\frac{\ii\omega m q\e{s}}{\gamma v c\e{s}^2}
    \int_0^\infty \frac{k_r \dd k_r}{2\pi} \;
    \frac{J_0(R k_r)}{k_r^2 + p^2(\omega)}
    \label{eq:dph0_intermediate}
\\
&= -\frac{\ii\omega m q\e{s}}{2\pi \gamma v c\e{s}^2} \,
    K_0[R p(\omega)] \, ,\label{eq:dph0}
\end{align}
where $J_n, K_n$ denote Bessel functions of the first and second kind, respectively, and we have introduced
\begin{equation}
\label{eq:definition_p_squared}
p^2(\omega) \equiv
\left( \frac{1}{v^2} - \frac{1}{c\e{s}^2} \right) \omega^2 \, .
\end{equation}
Note that $p(\omega)$ is real if the particle is subscalar ($v\leq c\e{s}$) and imaginary if the particle is superscalar ($v> c\e{s}$).

As for the second important term in Eq.~\eqref{eq:dEdt_Fourier}, we note that
$\widehat{\dph_{,r}}(\omega, R)
= \dd\widehat{\dph}(\omega, R)/\dd R
= \ii\omega^{-1} \dd\widehat{\dph_{,0}}(\omega, R)/\dd R$,
and use that $\dd K_0(x)/\dd x= -K_{1}(x)$ to find
\begin{equation} \label{eq:dphr}
\widehat{\dph_{,r}}(\omega, R)
= -\frac{m q\e{s}}{2\pi \gamma v c\e{s}^2} \,
    p(\omega) \, K_1[R p(\omega)] \, .
\end{equation}
Plugging Eqs.~\eqref{eq:dph0} and \eqref{eq:dphr} into Eq.~\eqref{eq:dEdt_Fourier}, we finally get the energy loss per unit distance to scalar waves
\begin{align}
\frac{\dd E}{\dd x}
&= \frac{1}{v} \frac{\dd E}{\dd t}
\\
&= -\frac{\ii \P}{4\pi^2} \l(\frac{m}{\gamma v}\r)^2
\int_{\mathbb{R}} \dd \omega \; \omega \l(\frac{q\e{s}}{c\e{s}^2}\r)^2 R p \, K_0^*(R p) K_1(R p) \, .
\label{eq:dEdx_intermediate}
\end{align}
Here, we keep $q\e{s}$ and $c\e{s}$ inside the integral since they are, in principle, frequency dependent.

\subsection{Scalar \texorpdfstring{$\check{\hbox{C}}$}{Ch}erenkov radiation}

Because of energy conservation, $\dd E/\dd x$ must not depend on the surface that is chosen to evaluate the flux of radiated energy; in particular, it should not depend on the radius~$R$ of the cylinder. We can thus evaluate $\dd E/\dd x$ in the limit $R\rightarrow\infty$.
For that purpose, we may substitute in Eq.~\eqref{eq:dEdx_intermediate} the asymptotic form of the Bessel functions~$K_n(z)$ for $|z|\gg 1$ and $\arg(z) < 3\pi/2$,
\begin{align}
K_n(R p) \sim \sqrt{\frac{\pi}{2 R p}} \, \ex{- R p} \, ,
\end{align}
to find, in the limit $R\rightarrow\infty$,
\begin{align}
\frac{\dd E}{\dd x}
&\sim -\frac{\ii \P}{8\pi} \l(\frac{m}{\gamma v}\r)^2
\int_{\mathbb{R}} \dd \omega \; \omega
\l(\frac{q\e{s}}{c\e{s}^2}\r)^2 \sqrt{\frac{p}{p^*}} \, \ex{- 2R \Re(p)} \, .
\end{align}

At that point, we must recall the definition~\eqref{eq:definition_p_squared} of $p^2$, which implies that $p(\omega)=\pm \omega \sqrt{1/v^2 - 1/c\e{s}^2}$. The $\pm$ sign translates an ambiguity that we omitted when going from Eq.~\eqref{eq:dph0_intermediate} to Eq.~\eqref{eq:dph0}. If the particle is subscalar, i.e., $v<c\e{s}$, then either $p>0$ in which case $\dd E/\dd x\to 0$ as $R\to\infty$; or $p<0$ in which case $\dd E/\dd x$ diverges along the imaginary axis, which is nonphysical. If the particle is superscalar, i.e.,  $v>c\e{s}$, then $p\propto\pm\ii\omega$ and hence
\begin{equation}
\label{eq:energy_change_ambiguity}
\frac{\dd E}{\dd x}
= \pm \frac{\P}{8\pi} \l(\frac{m}{\gamma v}\r)^2
\int_{\mathbb{R}} \dd \omega \; \omega
\l(\frac{q\e{s}}{c\e{s}^2}\r)^2 \, .
\end{equation}
Negative frequencies may be associated with scalar waves emitted from infinity and being absorbed by the particle, thereby increasing its energy, while positive frequencies may be associated with scalar waves emitted by the particle and radiated away. This (conventional) association implies that the negative sign should be chosen in Eq.~\eqref{eq:energy_change_ambiguity}. If we only consider the waves emitted by the particle ($\omega\geq 0$) and recall that the whole calculation is only valid above the IR cutoff and below the UV cutoff, we conclude that
\begin{empheq}[box=\fbox]{align}
\l(\frac{\dd E}{\dd x}\r)\e{rad}
=  -\frac{\P}{8\pi} \l(\frac{m}{\gamma v}\r)^2
    \int_{\Omega\e{IR}}^{\Omega\e{UV}} \dd \omega \; \omega
    \l(\frac{q\e{s}}{c\e{s}^2}\r)^2 . \label{eq:dEdx_integral}
\end{empheq}
Assuming for simplicity that the scalar speed $c\e{s}$ is constant over $\omega \in [\Omega\e{IR},\Omega\e{UV}]$ and that $\Omega\e{IR}\ll\Omega\e{UV}$, we find for $v>c\e{s}$,
\begin{align}\label{eq:dEdx_pheno}
\l(\frac{\dd E}{\dd x}\r)\e{rad}
= - \frac{\P (q\e{s} M\e{P})^2}{16 \pi c\e{s}^4} \l(\frac{\Omega\e{UV}}{M\e{P}}\r)^2 \l( \frac{m}{\gamma v}\r)^2\,.
\end{align}
We observe without surprise that the energy loss is proportional to $\P$ and to the scalar coupling constant squared $(q_s M\e{P})^2$. We also note that the relativistic Lorentz factor $\gamma^{-2}=1-v^2$ suppresses the energy loss for ultra-relativistic particles. We could expect that the radiated energy vanishes as $v\rightarrow 1$, because photons have a traceless energy-momentum tensor, thereby eliminating the source from the start. Perhaps more unexpected is the $c\e{s}^{-4}$ dependence, which highly enhances the energy loss if the scalar-wave speed is low.

In the k-essence case investigated here, we may replace $c\e{s}^2, q\e{s}, \P$ with their expressions \eqref{eq:cs2}, \eqref{eq:qs} and \eqref{eq:P} to get
\begin{align}\label{eq:dEdx_kessence}
\l(\frac{\dd E}{\dd x}\r)\e{rad}
=
-\frac{1}{32\pi \bG_4}
\frac{\bG_{4,\ph}^2 \bG_4^{-1}}{\bG_{2,X}+3\bG_{4,\ph}^2\bG_4^{-1}}
\l( \frac{\Omega\e{UV}}{M\e{P}} \r)^2
\l( \frac{m}{\gamma v} \r)^2 \,.
\end{align}
Interestingly, the dependence on $\bG_{2,XX}$ disappears from the expression for the energy loss; it however remains important for the condition of validity $v>c\e{s}$. This cancellation is due to the fact that $q\e{s}$ and $c\e{s}^2$ have the same denominator and that there is no other dependence of the energy loss per unit distance on $\bG_{2,XX}$. In fact, when $\bG_{2,XX} \to 0$, the condition $v>c\e{s}$ forces $v\to 1$ and the Lorentz factor $\gamma^{-2}= 1- v^2$ drives the energy loss per unit distance to vanish, as one may expect. Note that the prefactor $ (3 + \bG_{2,X} \bG_4 \bG_{4,\ph}^{-2})^{-1}$ which depends on the functional form of $\bG_{2}$ and $\bG_4$ cannot be large. In fact, stability requirements ($\bG_{2,X} + 2 X \bG_{2,XX}>0$ together with $c\e{s}^2>0$) imply $\bG_{2,X} + \bG_{4,\ph}^2 \bG_4^{-1}>0$ and $\bG_{4}>0$, which bound this dimensionless coupling to belong to $[0, 1/3]$. The situation where the conformal coupling is arbitrarly small (i.e., $\bG_4 \to 0$) could drive the proportionality factor to be large, however, we do not expect it to deviate too much from unity in dark-energy-motivated scenarios~\cite{LISACosmologyWorkingGroup:2019mwx}. 

\section{Discussion}\label{sec:discussion}

In the previous section, we have computed the energy loss~\eqref{eq:dEdx_integral} to scalar waves from a point-like particle moving faster than those. We have found that in the nearly constant velocity case, it scales with the square of the UV cutoff, which makes it impossible to ignore the effective-field theory nature of the conformally coupled k-essence action. If viewed as a covariant theory valid at all scales, or as an EFT valid all the way up to the Planck scale, as may be the case if motivated by quantum-gravity perspectives, then the energy loss to scalar waves may diverge or become dramatically large and extremely tight constraints could be placed on a combination of $\bG_2$ and $\bG_4$ from the observation of high-energy cosmic rays of extragalactic origin. In the next two subsections, we consider two situations of interest, conformally coupled k-essence and a phenomenological approach based on the parameters that appear in the equation of motion \eqref{eq:EoM}.

\subsection{Conformally coupled k-essence}

To get an order of magnitude for the energy loss of a proton to scalar waves coming from cosmological distances, we use the fact that the $\bG_4(\ph)$ and $\bG_{2}(\ph,X)$-dependent prefactor in Eq.~\eqref{eq:dEdx_kessence} is expected to be of order unity \cite{Traykova:2021hbr}, which results in the energy loss per unit distance
\begin{align}\label{eq:dEdx57}
\l(\frac{\dd E}{\dd x}\r)\e{rad}
\sim 
-\frac{1}{32\pi}
\l( \frac{\Omega\e{UV}}{M\e{P}} \r)^2
\l( \frac{m}{\gamma v} \r)^2 \,.
\end{align} 
In the limit of small fractional energy loss, denoting the initial energy of the moving particle by $E\e{i} = m \gamma(v\e{i})$, then the former may be written as
\begin{align}
\l|\frac{\Delta E}{E\e{i}}\r|
\equiv
\l| \frac{E(x) - E\e{i}}{E\e{i}} \r|
\sim \frac{x}{\ell}\,,
\end{align}
where we defined a characteristic length $\ell$ beyond which particles effectively slow down to the speed of the scalar waves
\begin{align}
\ell
\equiv
\frac{16\pi \gamma^3 v^2}{m}
\l( \frac{M\e{P}}{\Omega\e{UV}} \r)^2\,. \label{eq:x_o}
\end{align}
As an illustrative example, we fix $x=1$ Gpc, $v=0.9$, for a proton mass $m=m\e{p}$ and plot the fractional energy loss as a function of $\Omega\e{UV}$ in Fig.~\ref{fig:DeltaEoverE}. The energy loss to scalar waves is mostly negligible in this situation, unless the UV cutoff satisfies $\Omega\e{UV}> 10^{-20} M\e{P}$. Recall that for DE motivated scenarios, the UV cutoff can be of the order of $(M\e{P} H_0^2)^{1/3} \simeq 10^{-40} M\e{P}$ \cite{deRham:2018red}. It turns out that for nonrelativistic scalar speeds, the scale at which the EFT goes out of control can be much lower than the UV cutoff by a factor of $c\e{s}^n$ with $n>0$ \cite{deRham:2017aoj}. The value of $n$ depends on the specific irrelevant operators that are considered. This implies that our estimate of the fractional energy loss can be considered as an upper bound on the amount of energy that is actually lost to scalar waves.\footnote{We thank the anonymous referee for pointing this out.} For quasiluminal scalar waves, this $c\e{s}^n$ suppression does not change the order of magnitude of the radiated energy. In fact, for a UV cutoff equal to the Planck mass, any $c\e{s}^n\geq 10^{-20}$ implies that the energy loss is dramatically large, in the sense that Eq.\,\eqref{eq:dEdx57} can not be integrated assuming the limit of small fractional energy loss. This is such that cosmic rays may lose a significant portion of their energy over cosmological distances. In the following, we consider a more generic phenomenological approach, where $c\e{s}$ bears no relationship to $\P$ and $q\e{s}$, which allows the prefactor in \eqref{eq:dEdx_pheno} to become large in case of low scalar phase velocity and for fixed UV cutoffs. 

\subsection{Phenomenological approach}
In this subsection, we consider the case where there is enough freedom for $q\e{s}$, $c\e{s}$ and $\P$ to be independent. We consider the situation where $\P\sim M\e{P}^2/2$, $q\e{s}\sim M\e{P}^{-2}$ and yet $c\e{s}\ll 1$.  This may allow for the energy loss per unit distance \eqref{eq:dEdx_pheno} to become large. This situation may be realized for models with enough freedom, for instance where a cubic Galileon term is present in the action. In this case, we find that 
\begin{align}
\l(\frac{\dd E}{\dd x}\r)\e{rad}
\sim 
-\frac{1}{32\pi c\e{s}^4}
\l( \frac{\Omega\e{UV}}{M\e{P}} \r)^2
\l( \frac{m}{\gamma v} \r)^2 \,.
\end{align} 
We note that for $v<c\e{s}$, there is no radiated energy and for $v\to 1$, the energy loss also vanishes. Therefore, since $(\dd E/\dd x)\e{rad}(v)$ is a continuous function of $v\in[c\e{s},1]$, by the extreme value theorem, there exists a minimum and maximum energy loss. The energy loss is maximal for $v \in [c\e{s},1]$ and bounded from below by zero. If the spectrum of high-energy particles is known, for example, to be a decaying power law, the signature of this scalar $\check{\hbox{C}}$erenkov radiation would be a well in the spectrum, followed by a compensating peak at the energy corresponding to $v=c\e{s}$.

\begin{figure}
    \centering
    \includegraphics[width=\columnwidth]{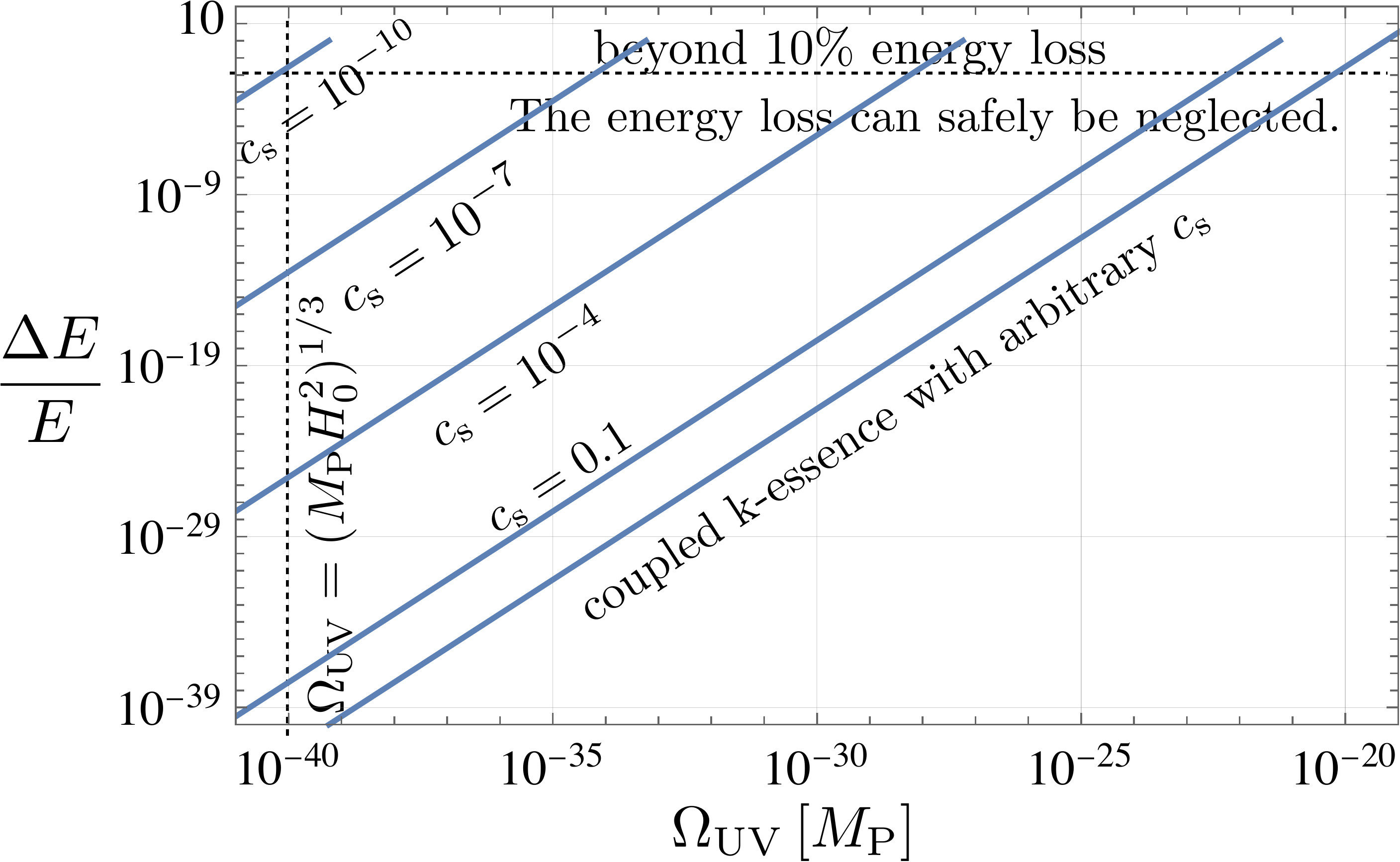}
    \caption{Energy loss to scalar waves for a proton traveling $1$ Gpc at $v = 0.9 c$ in a Universe where scalar waves propagate at $c\e{s} \in \{10^{-1},10^{-4},10^{-7},10^{-10}\}$ as a function of the UV cutoff $\Omega\e{UV}$. For $c\e{s}=10^{-4}$, the energy loss gets above $10\%$ and becomes rapidly large beyond that for UV cutoffs of the order of $\Omega\e{UV}\geq 10^{-27} M\e{P}$. For any cutoff below that, the energy loss is negligible. In particular, for Horndeski theories of cosmological interests, $\Omega\e{UV}$ can be of the order of $c\e{s}^n(M\e{P} H_0^2)^{1/3} \sim c\e{s}^n 10^{-40}M\e{P}$, with $n>0$ for which the energy loss is completely negligible. The blue lines depict phenomenological models with different values of $c\e{s}$, including k-essence with arbitrary $c\e{s}<v$. Their intersection with $\Omega\e{UV} = (M\e{P}H_0^2)^{1/3}$ can be considered as upper bounds on the fractional energy loss.}
    \label{fig:DeltaEoverE}
\end{figure}

In the limit of small fractional energy loss,
\begin{align}
\l|\frac{\Delta E}{E\e{i}}\r| \sim \frac{x}{\ell}\,,
\qquad
\ell
\equiv
\frac{16\pi \gamma^3 v^2 c\e{s}^4}{m}
\l( \frac{M\e{P}}{\Omega\e{UV}} \r)^2\,.
\end{align}
For the same illustrative example as in the previous susbsection,  fixing $x=1$ Gpc, $v=0.9$, for a proton mass $m=m\e{p}$, we plot the fractional energy loss as a function of $\Omega\e{UV}$ for different values of $c\e{s}\in\{10^{-10},\, 10^{-7},\,10^{-4},\,10^{-1} \}$ in Fig.\,\ref{fig:DeltaEoverE}.
The energy loss can be safely neglected as long as it represents less than, say $10$\% of the initial energy. For $c\e{s}= 10^{-4}$, this corresponds to a UV cutoffs satisfying $\Omega\e{UV}\lesssim 10^{-27} M\e{P} \sim 10^{-9} m\e{p}$. If the scalar-tensor theory is aimed at explaining cosmic acceleration, then the UV cutoff can be of the order $c\e{s}^n (M\e{P}H_0^2)^{1/3} \simeq c\e{s}^n 10^{-40} M\e{P}$, for which the energy loss to scalar waves is totally negligible.
\begin{figure}
    \centering
    \includegraphics[width=\columnwidth]{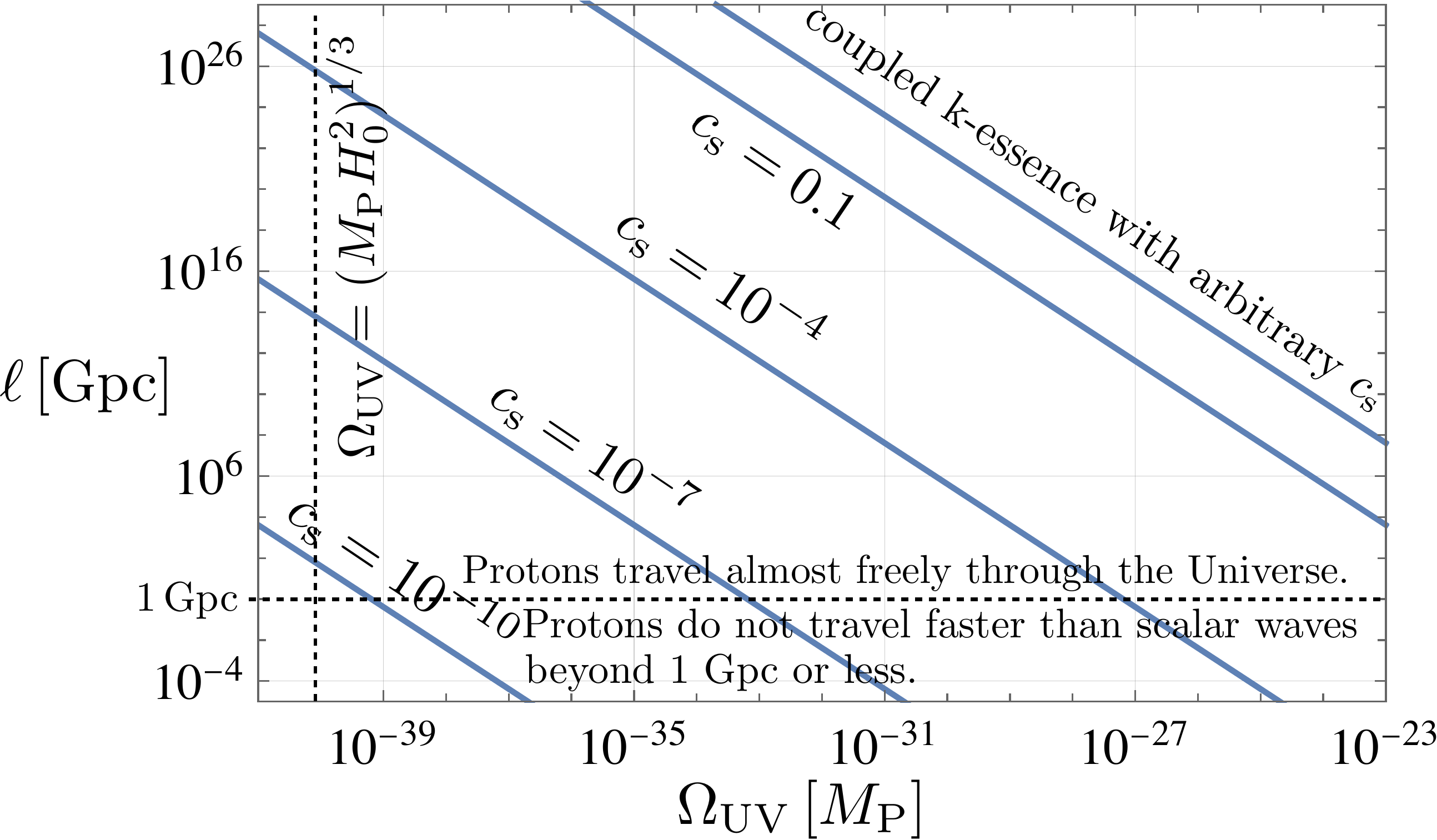}
    \caption{Characteristic length for a proton traveling at $v = 0.9$ in a Universe where scalar waves propagate at $c\e{s}\in \{10^{-10},10^{-7},10^{-4},10^{-1}\}$, as a function of the UV cutoff $\Omega\e{UV}$. For $c\e{s}=10^{-4}$ and UV cutoffs above the order of $\Omega\e{UV}\geq 10^{-28} M\e{P}$, the characteristic length goes below $1$~Gpc, slowing down efficiently protons to the speed of scalar waves. For any cutoff below that, protons travel freely through the Universe. In particular, for Horndeski theories of cosmological interests, $\Omega\e{UV}$ can be of order $c\e{s}^n (M\e{P} H_0^2)^{1/3} \sim c\e{s}^n 10^{-40}M\e{P}$, with $n>0$, the characteristic length is much larger than the size of the observable Universe, except if the sound speed is as small as $c\e{s}\leq 10^{-10}$ and $n$ is reasonably small. The blue lines depict phenomenological models with different values of $c\e{s}$, including k-essence with arbitrary $c\e{s}<v$.}
    \label{fig:Characteristic_Length}
\end{figure}
We also plot the characteristic length $\ell$ as a function of the UV cutoff for different values of $c\e{s}$ in Fig.~\ref{fig:Characteristic_Length}. For $c\e{s}=10^{-4}$, if the UV cutoff is above $10^{-28}M\e{P}$, then the proton slows down to the scalar speed in less than $1$ Gpc. For DE-motivated UV cutoffs, the same proton travels freely through the Universe. Thus, even if massive particles interact with a nonminimally coupled subluminal scalar field, dark energy EFT considerations prevent the energy loss from being significant. This result may change in other contexts. For example, if the scalar phase velocity is much smaller, of the order of $c\e{s}=10^{-10}$, than the constraint can become significant because of the $c\e{s}^4$ dependence in Eq.~\eqref{eq:x_o}. However, this $c\e{s}^4$ amplification may be altered by a suppression of the effective UV cutoff for non-relativistic scalar fields by an extra factor $c\e{s}^n$ \cite{deRham:2017aoj}. 

Finally, let us mention the case where the k-essence field is motivated as a dark-matter candidate~\cite{Brax:2020tuk}. In this case, tight constraints from collider experiments should be implemented on $G_4$, but the sound speed can be motivated to be small. Note that screening mechanisms could in principle allow to evade laboratory tests of $G_4$, in which case, the constraint that we derived in this work could be competitive. In this sense, scalar $\check{\hbox{C}}$erenkov radiation could be used as a complementary constraint on the existence of this kind of scalar field. We leave this perspective to future work.

\section{Conclusions}\label{sec:conclusions}
In this work, we studied for the first time the energy loss of a massive particle to scalar waves in a conformally coupled k-essence theory. We derived an equivalent of the Frank-Tamm formula for scalar $\check{\hbox{C}}$erenkov waves valid when the massive particle travels faster than the phase velocity of scalar waves. We gave the explicit dependence on the free functions of conformally coupled k-essence and also a more generic dependence on the scalar phase velocity $c\e{s}$ and effective scalar charge $q\e{s}$. We found that the energy loss per unit distance scales with the square of the UV cutoff of the EFT of gravity, which makes it impossible to ignore the EFT nature of a conformally coupled k-essence theory. We then showed that the energy loss is negligible for conformally coupled k-essence as long as the UV cutoff is lower than $10^{-20} M\e{P}$. 

For dark-energy motivated EFTs, the UV cutoff can be as low as $\Omega\e{UV}\sim (M\e{P}H_0^2)^{1/3}$, which translates to a completely negligible radiated energy in the form of scalar waves for a proton travelling through the Universe. In contrast, for $\Omega\e{UV} \sim M\e{P}$, as long as $c\e{s}^n\geq 10^{-20}$, the energy loss is dramatically large and the speed of cosmic rays would be changed during the propagation over cosmological distances. This is likely to result in a change in the distribution of speeds for high-energy cosmic rays of cosmological origin.

For phenomenological models with a low sound speed $c\e{s}$, the energy loss per unit distance is amplified by a factor $c\e{s}^{-4}$, which can become significant for lower UV cutoffs, although this effect may be compensated by a loss of the trustability of the EFT at energy scales lower by a factor $c\e{s}^n$.  Nevertheless, if one adopts a UV cutoff of the order of $(M\e{P}H_0^2)^{1/3}$ as may be motivated to explain cosmic acceleration, the scalar phase velocity must be as low as $c\e{s}\sim 10^{-10}$ for the effect to be significant on a proton traveling cosmological distances.

\section{Acknowledgements}

C.D.\,thanks Farbod Hassani and Emilio Bellini for useful discussions. The authors would like to thank Tobias Mistele for spotting a typo in the first version of this work. C.D.\,and L.L.\,were supported by a Swiss National Science Foundation (SNSF) Professorship grant (Nos.\,170547 \& 202671). C.D.\,further acknowledges support by the Department of Theoretical Physics, University of Geneva. P.F.\,received the support of a fellowship from ``la Caixa'' Foundation (ID 100010434). The fellowship code is LCF/BQ/PI19/11690018.

\bibliographystyle{unsrt}
\bibliography{references}

\end{document}

%% file: packages.tex
\usepackage[english]{babel}

\usepackage[utf8x]{inputenc}
\usepackage[T1]{fontenc}
\usepackage{ucs}
\usepackage{microtype}
\usepackage{newtxtext,newtxmath}
\usepackage{color}
\usepackage{lettrine}

\usepackage{url}
\usepackage{hyperref}

\usepackage{xspace}
\usepackage{lastpage}
\usepackage{stmaryrd}

\usepackage{graphicx}
\usepackage{empheq}
\usepackage{ifthen}
\usepackage{tensor}
\usepackage{paralist}
\usepackage{float}

\usepackage{slashed}
\usepackage{amsmath}
\usepackage{amsfonts}
\usepackage{amssymb}
\usepackage{mathrsfs}
\usepackage{extarrows}
\usepackage{verbatim}
\usepackage{upgreek}
\usepackage{wasysym}

\usepackage{natbib}